\documentstyle[prl,aps,multicol,psfig]{revtex}
\begin{document}
\draft
\title{Metal-insulator transition in a 2D electron gas: Equivalence
of two approaches for determining the critical point}
\author{A.~A. Shashkin$^*$ and S.~V. Kravchenko}
\address{Physics Department, Northeastern University, Boston,
Massachusetts 02115}
\author{T.~M. Klapwijk}
\address{Department of Applied Physics, Delft University of
Technology, 2628 CJ Delft, The Netherlands}
\date{\today}
\maketitle

\begin{abstract}
The critical electron density for the metal-insulator transition in a
two-dimensional electron gas can be determined by two distinct
methods: (i)~a sign change of the temperature derivative of the
resistance, and (ii)~vanishing activation energy and vanishing
nonlinearity of current-voltage characteristics as extrapolated from
the insulating side. We find that in zero magnetic field (but not in
the presence of a parallel magnetic field), both methods give
equivalent results, adding support to the existence of a true
zero-field metal-insulator transition.
\end{abstract}
\begin{multicols}{2}

The observation of a metal-insulator transition (MIT) in
two-dimensional (2D) electron and hole systems (see a review
\cite{abrahams00} and references therein) challenges the scaling
theory of localization which predicts all electron states to be
localized in an infinite disordered 2D system at zero temperature and
in zero magnetic field \cite{abrahams79}. In fact, the predicted
localization of all carriers is beyond experimental verification
since in real 2D systems, it can break down because of any
perturbation such as a finite temperature, finite system size,
magnetic impurities, etc. To estimate the limiting behavior at zero
temperature experimentally, one can follow two principal approaches:
extrapolate the temperature dependence of the resistance to $T=0$, or
analyze temperature-independent characteristics ({\it e.g.}, the
correlation length). Moreover, comparison of distinct criteria is the
experimental test to find out whether or not a true MIT exists. The
resistance of high-mobility silicon and $p$-GaAs samples is ``flat''
at a certain carrier density in a wide range of temperatures
\cite{sarachik99,kravchenko00,mills}, and it is therefore tempting to
extrapolate it to $T=0$ and identify the corresponding electron
density with the critical density for the metal-insulator transition.
However, in some of the suggested explanations for the MIT (see, {\it
e.g.}, Ref.~\cite{meir99}), it was stated that within their models,
the ``critical'' curve is not flat. Therefore, to verify whether or
not the temperature-independent resistance corresponds to the
critical density, an independent determination of the critical point
is necessary. This is why in this paper, we compare the critical
density obtained by a finite-temperature criterion with that obtained
by a {\em temperature-independent} one. Equivalent results from these
two distinct methods would justify the extrapolation of the flat
resistivity curve to $T=0$ and strongly support the existence of a
true MIT in zero magnetic field.

Methods for determining the MIT point have been described in previous
publications. The first criterion is a change in sign of the
temperature derivative of the resistivity, ${\rm d}\rho/{\rm d}T$
(see, {\it e.g.}, Ref.~\cite{kravchenko00}). A positive (negative)
sign of the derivative at the lowest achievable temperatures is
empirically associated with a metallic (insulating) phase. The second
criterion is based on a vanishing activation energy, $E_a$, combined
with a vanishing nonlinearity (threshold voltage, $V_c$) of
current-voltage ($I-V$) characteristics when extrapolated from the
insulating phase \cite{shashkin94,note}. The activation energy and
the threshold voltage are connected via the localization length which
is {\it temperature-independent}. While the derivative method deals
with the vicinity of the MIT in which the dependence $\rho(T)$ is
relatively weak, the $I-V$ method is related to the {\em insulating
phase} with exponential $\rho(T)$. These two methods have not been
applied simultaneously to the 2D MIT.

In this Letter, we compare these two independent criteria for
determining the metal-insulator transition point in a 2D electron
system in silicon metal-oxide-semiconductor field-effect transistors
(MOSFETs). We report that in zero magnetic field, both methods yield
the same critical electron density $n_c(0)$. Since one of the methods
is temperature-independent, this equivalence supports the existence
of a true $T=0$ MIT in zero magnetic field. In contrast, in high
parallel magnetic fields, where the 2D electrons are fully
spin-polarized \cite{Ok}, only the $I-V$ method can be used, and it
yields the critical density $\approx1.5\, n_c(0)$; the derivative
criterion does not yield a critical point. This makes uncertain the
existence of a true metal-insulator transition in a system of
spin-polarized electrons.

Measurements were performed on high-mobility silicon MOSFETs
($\mu_{\text{peak}}=2.4\times 10^4$~cm$^2$/Vs at 4.2~K) similar to
those previously used in Ref.~\cite{kravchenko00}. Data were taken by
a four-terminal {\it dc} technique using an electrometer with
symmetric inputs.

The resistivity as a function of temperature in zero magnetic field
is shown in Fig.~\ref{1}(a) for several electron
\end{multicols}
\vbox{
\vspace{12.6mm}
\hbox{
\psfig{file=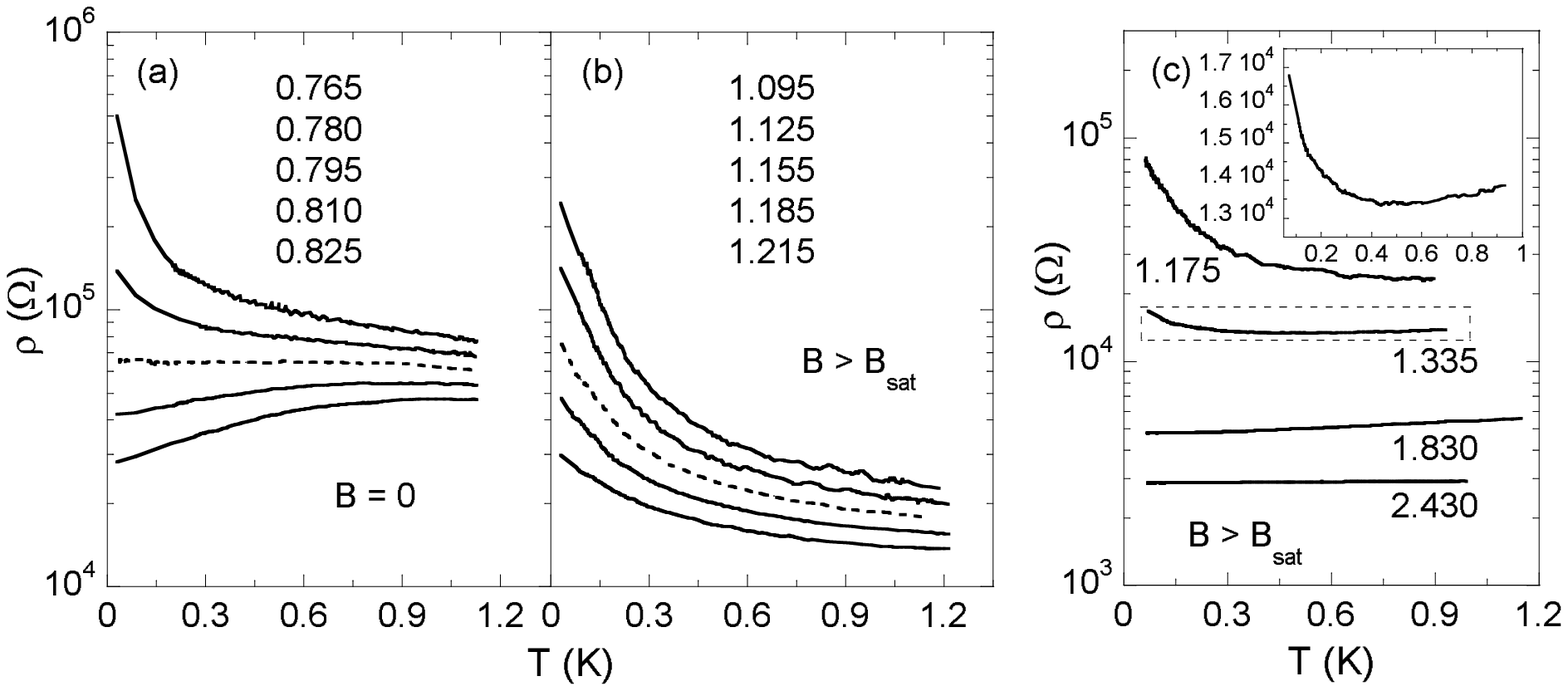,width=5.4in,bbllx=-.8in,bblly=1.25in,bburx=7.25in,bbury=9.5in,angle=0}
}
\vspace{-3.6in}
\hbox{
\hspace{-0.1in}
\refstepcounter{figure}
\parbox[b]{7.0in}{\baselineskip=12pt \egtrm FIG.~\thefigure.
Temperature dependence of the resistivity of a Si MOSFET at different
electron densities near the MIT in zero magnetic field (a), and in
parallel magnetic fields of 4~Tesla (b) and 10~Tesla (c). The
electron densities are indicated in units of $10^{11}$~cm$^{-2}$.
The inset in (c) shows a close-up view of the curve marked by the
rectangle.\vspace{0.05in}
}
\label{1}
}
}
\begin{multicols}{2}
\noindent
densities on both
sides of the metal-insulator transition. The resistivity of the
middle (dashed) curve shows virtually no temperature dependence over
a wide range of temperatures; this curve separates those with
positive and negative ${\rm d}\rho/{\rm d}T$ nearly symmetrically (at
temperatures above 0.2~K) as reported earlier \cite{simonian97}. The
existence of such a $T$-independent curve at temperatures down to
30~mK clearly shows that the logarithmic corrections to the
resistance (which are expected to be very strong in 2D systems with
resistivity $\gtrsim h/e^2$ \cite{abrahams79}) are absent in this
system in zero magnetic field, see also
Refs.~\cite{kravchenko00,mills}. Assuming that the middle curve
remains flat down to $T=0$, we obtain the MIT critical point at
$n_s=0.795\times 10^{11}$~cm$^{-2}$ which corresponds to a
resistivity $\rho\approx 3h/e^2$, as in other experiments on Si
MOSFETs \cite{abrahams00}. We designate the corresponding electron
density as the critical density, $n_{c1}$.

The MIT point has also been determined by studying the behavior of
nonlinear $I-V$ characteristics on the insulating side of the
transition. Deep in the insulating state ($n_s<n_{c1}$), a typical
low-temperature $I-V$ curve is close to a step-like function: the
voltage rises abruptly at low current and then saturates, as shown in
Fig.~\ref{2}(a). The magnitude of the step is $2\, V_c$. (At higher
temperatures the curve becomes less sharp, yet the threshold voltage,
$V_c$, remains practically unchanged.) Closer to the MIT, the $I-V$
curves still show a nonlinear step-like behavior provided that
$n_s<n_{c1}$ (see the curve in Fig.~\ref{2}(b) corresponding to
$n_s=0.743\times 10^{11}$~cm$^{-2}$). Exactly at
$n_s=n_{c1}=0.795\times 10^{11}$~cm$^{-2}$, the $I-V$ curve is
strictly linear (Fig.~\ref{2}(b)); at $n_s>n_{c1}$, it becomes {\em
superlinear} which is characteristic of the metallic state
\cite{efield}. Figure~\ref{3} (closed circles) shows that the square
root of the threshold voltage is a linear function of electron
density (discussed below). Extrapolation of the $V_c^{1/2}(n_s)$
dependence to zero threshold value yields the critical electron
density $n_{c2}$. Note that $n_{c2}$ is equal to $n_{c1}$ with high
accuracy.

Not too deep on the insulating side of the transition, and at not too
low temperatures, the resistance has an activated form
\cite{shashkin94,diorio92}, as shown in the inset of Fig.~\ref{2}(a)
\cite{scaling}. Such a form of temperature dependence was interpreted
in Ref.~\cite{adkins76} as a result of the thermal activation of
carriers to the mobility edge, $E_c$; in this case, the activation
energy is $E_a=E_c-E_F$. Since in the insulating regime, the $I-V$
curves are strongly nonlinear, we determine the resistivity from
${\rm d}V/{\rm d}I$ in the linear interval of $I-V$ curves, {\it
i.e.}, at $I\rightarrow0$. Figure~\ref{3} shows $E_a$ as a function
of the electron density (open circles). In previous detailed studies
\cite{shashkin94}, this dependence was found to be linear near the
mobility edge \cite{away}. The present data can also be approximated
by a linear function which yields, within the experimental
uncertainty, the same critical electron density $n_{c2}$ as the
square root of the threshold voltage. The reciprocal slope of
$E_a(n_s)$, $D^*$, can be interpreted \cite{adkins76} as the
thermodynamic density of states near the transition point.

The threshold behavior of the $I-V$ curves has been explained within
the concept of the breakdown in the insulating phase
\cite{shashkin94,nonlin}. Here we simply outline this concept. The
breakdown occurs when the localized electrons at the Fermi level gain
enough energy to reach the mobility edge in an electric field,
$V_c/d$, over a distance given by the localization length, $L$:
$eV_cL/d=E_c-E_F$, where $d$ is the distance between the potential
probes. The values $E_a$ and $V_c$ are related through the
localization length which is temperature-independent and diverges
near the transition as $L(E_F)\propto (E_c-E_F)^{-s}$ with exponent
$s$ close to unity \cite{shashkin94}. This corresponds to a linear
dependence $V_c^{1/2}(n_s)$ near the MIT, as seen in Fig.~\ref{3}.

We stress that in zero magnetic field, both methods --- the one based
on extrapolation of $\rho(T)$ to zero temperature and the other based
on the behavior of the temperature-independent localization length
--- give the same critical electron density. This adds confidence
that
\vbox{
\vspace{-5mm}
\hbox{
\hspace{0in}
\psfig{file=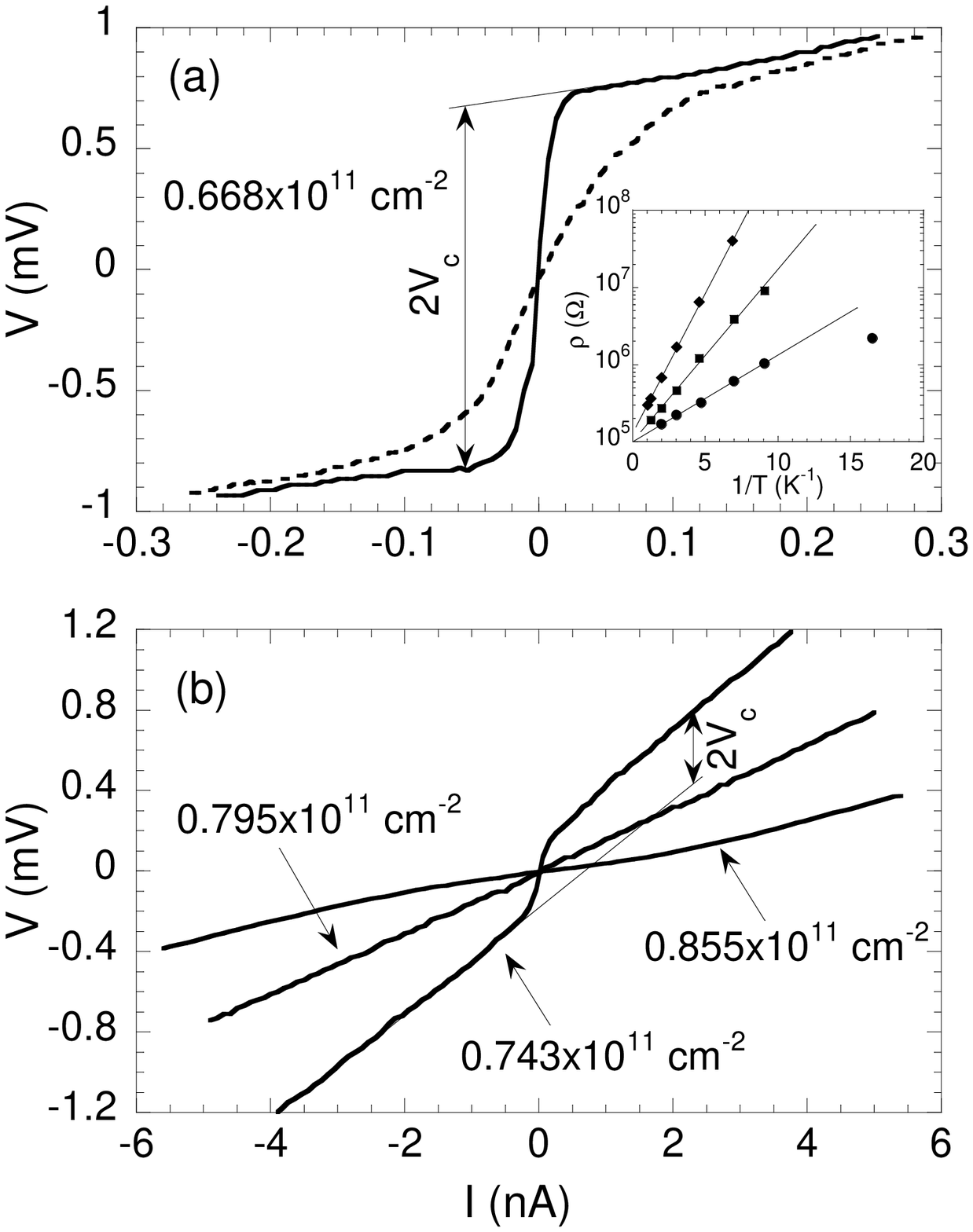,width=2.8in,bbllx=.5in,bblly=1.25in,bburx=7.25in,bbury=9.5in,angle=0}
}
\vspace{0.1in}
\hbox{
\hspace{-0.15in}
\refstepcounter{figure}
\parbox[b]{3.4in}{\baselineskip=12pt \egtrm FIG.~\thefigure.
Current-voltage characteristics at a temperature of $\approx 30$~mK
in zero magnetic field. (The aspect ratio is equal to 2.4 for this
sample.) In case (a), the $I-V$ curve obtained at a higher
temperature (211~mK; dashed line) is also shown for comparison; note
that the threshold voltage is practically temperature-independent. An
Arrhenius plot of the resistivity in the insulating phase is
displayed in the inset for the following values of $B$ and $n_s$:
0~T, $0.741\times10^{11}$~cm$^{-2}$ (circles); 1~T,
$0.810\times10^{11}$~cm$^{-2}$ (squares); 6~T,
$0.870\times10^{11}$~cm$^{-2}$ (diamonds).\vspace{0in}
}
\label{2}
}
}
\vbox{
\vspace{-10mm}
\hbox{
\hspace{2mm}
\psfig{file=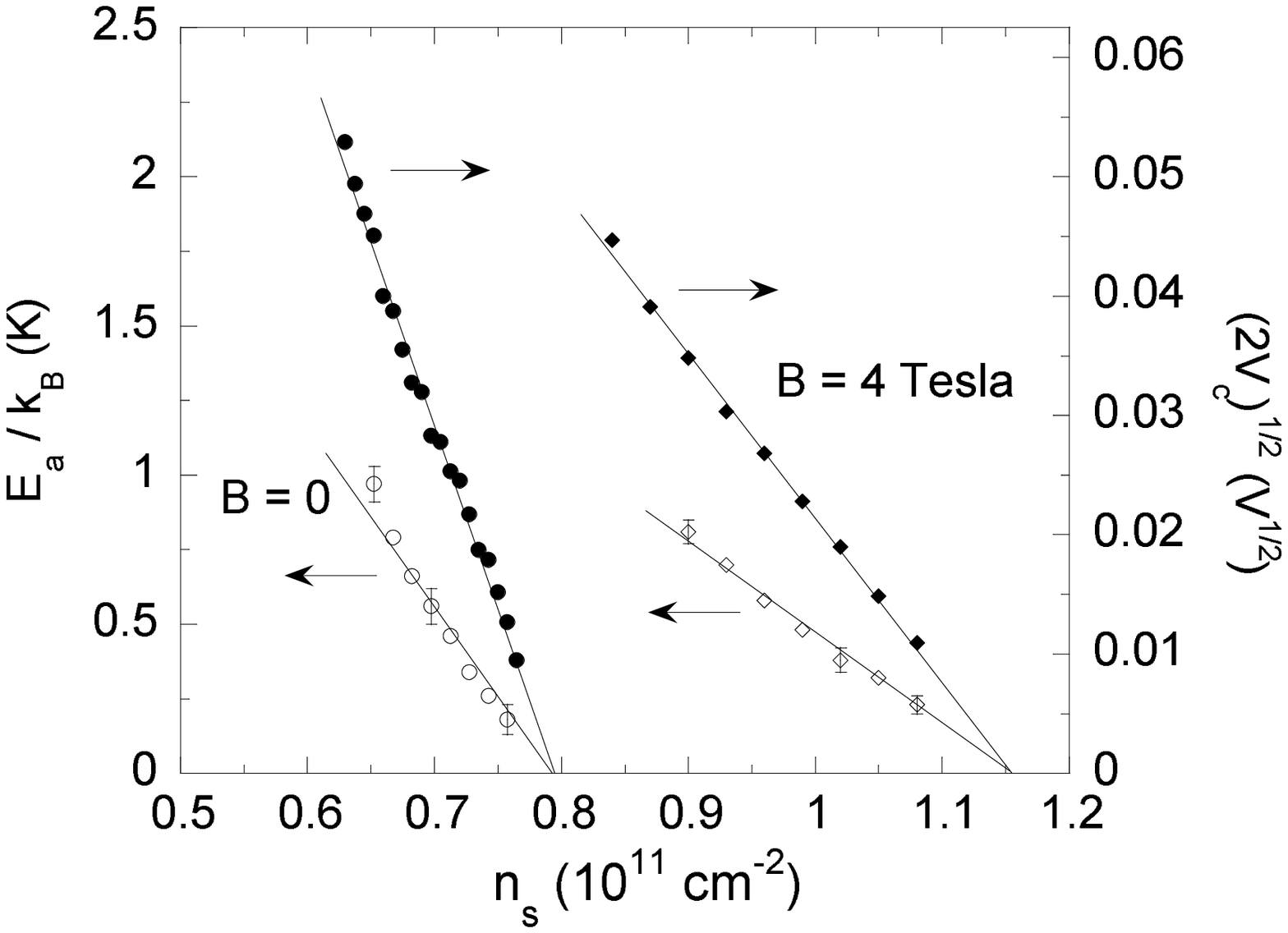,width=2.6in,bbllx=.5in,bblly=1.25in,bburx=7.25in,bbury=9.5in,angle=0}
}
\vspace{-0.6in}
\hbox{
\hspace{-0.15in}
\refstepcounter{figure}
\parbox[b]{3.4in}{\baselineskip=12pt \egtrm FIG.~\thefigure.
Activation energy and square root of the threshold voltage as a
function of electron density in zero magnetic field (circles) and in
a parallel magnetic field of 4~Tesla (diamonds).\vspace{0.20in}
}
\label{3}
}
}
the curve with zero derivative ${\rm d}\rho/{\rm d}T$ will
remain ``flat'' (or at least will retain finite resistivity value)
down to zero temperature.

It is interesting to compare the $B=0$ case with that in the presence
of a parallel magnetic field. With increasing parallel field, $B$,
the MIT point $n_{c2}$, determined from the vanishing nonlinearity
and activation energy, shifts to higher electron densities,
saturating above a critical field, $B_{\text{sat}}$, at a constant
value which is approximately 1.5 times higher than that in zero field
(see dots in Fig.~\ref{4}). A similar suppression of the metallic
behavior was observed using a resistance cut-off criterion at the
level on the order of $h/e^2$ \cite{dol92}. As was shown in
Ref.~\cite{Ok}, in the metallic phase the saturation of the
resistance with parallel field signals the onset of full spin
polarization of the 2D electrons. Hence, one expects that the 2D
system is spin-polarized at $B>B_{\text{sat}}$, and that the observed
phase boundary shift is a spin effect.

Before discussing the temperature dependence of the resistivity at
$B>B_{\text{sat}}$, we note that one cannot assume that the metallic
phase is necessarily strictly characterized by positive ${\rm
d}\rho/{\rm d}T$ \cite{rem}: one may have a weakly $T$-dependent
$\rho(T)$ with ${\rm d}\rho/{\rm d}T<0$ and still have a finite
resistivity at $T=0$. The $I-V$ method yields the electron density
$n_{c2}$ at which the {\it exponential} divergence of the resistivity
ends, although in principle ${\rm d}\rho/{\rm d}T$ may remain
negative at this density.

In Fig.~\ref{1}(b), we show the temperature dependence of the
resistivity in a parallel magnetic field high enough to cause full
spin polarization ($B=4$~Tesla). The middle curve corresponds to the
critical electron density, $n_{c2}(B>B_{\text{sat}})$, determined by
the method of vanishing nonlinearity and activation energy (as shown
in Fig.~\ref{3} by diamonds). In sharp contrast with the $B=0$
situation, not only are the $\rho(T)$ curves in the field
non-symmetric about the middle curve, but all of them have negative
``insulating-like'' derivatives ${\rm d}\rho/{\rm d}T<0$ in the
entire temperature range, although the values of the resistivity are
comparable to those in the $B=0$ case. Moreover, in a strong parallel
magnetic field, there is no temperature-independent $\rho(T)$ curve
at any electron density: as shown in Fig.~\ref{1}(c), the curve at
the considerably higher density $n_s=1.335\times 10^{11}$~cm$^{-2}$
compared to $n_{c2}=1.155\times 10^{11}$~cm$^{-2}$, which could be
approximately identified as a flat one in the temperature range used,
changes its slope from weakly-metallic at $T\gtrsim0.5$~K to
weakly-insulating at lower temperatures. The metallic behavior of the
resistance as a function of temperature, seen at yet higher electron
densities in a parallel magnetic field (Fig.~\ref{1}(c)), is much
weaker than in the absence of field. We therefore conclude that the
derivative method does not yield a critical density for the
spin-polarized 2D system. Its failure leaves uncertain the existence
of a true metal-insulator transition in a parallel magnetic field.

Some theories claim that spin-polarized and unpolarized states are
very similar \cite{degeneracy}. The authors of Ref.~\cite{GD}
considered the temperature-dependent screening of a random potential
and predicted metallic (${\rm d}\rho/{\rm d}T>0$) temperature
dependences of the resistivity for both polarized and unpolarized
states. Therefore, one might expect a more or less analogous behavior
of $\rho(T)$ curves around the transition in the two cases. However,
this is in contradiction with the experiment: while in zero magnetic
\vbox{
\vspace{-18.5mm}
\hbox{
\hspace{3mm}
\psfig{file=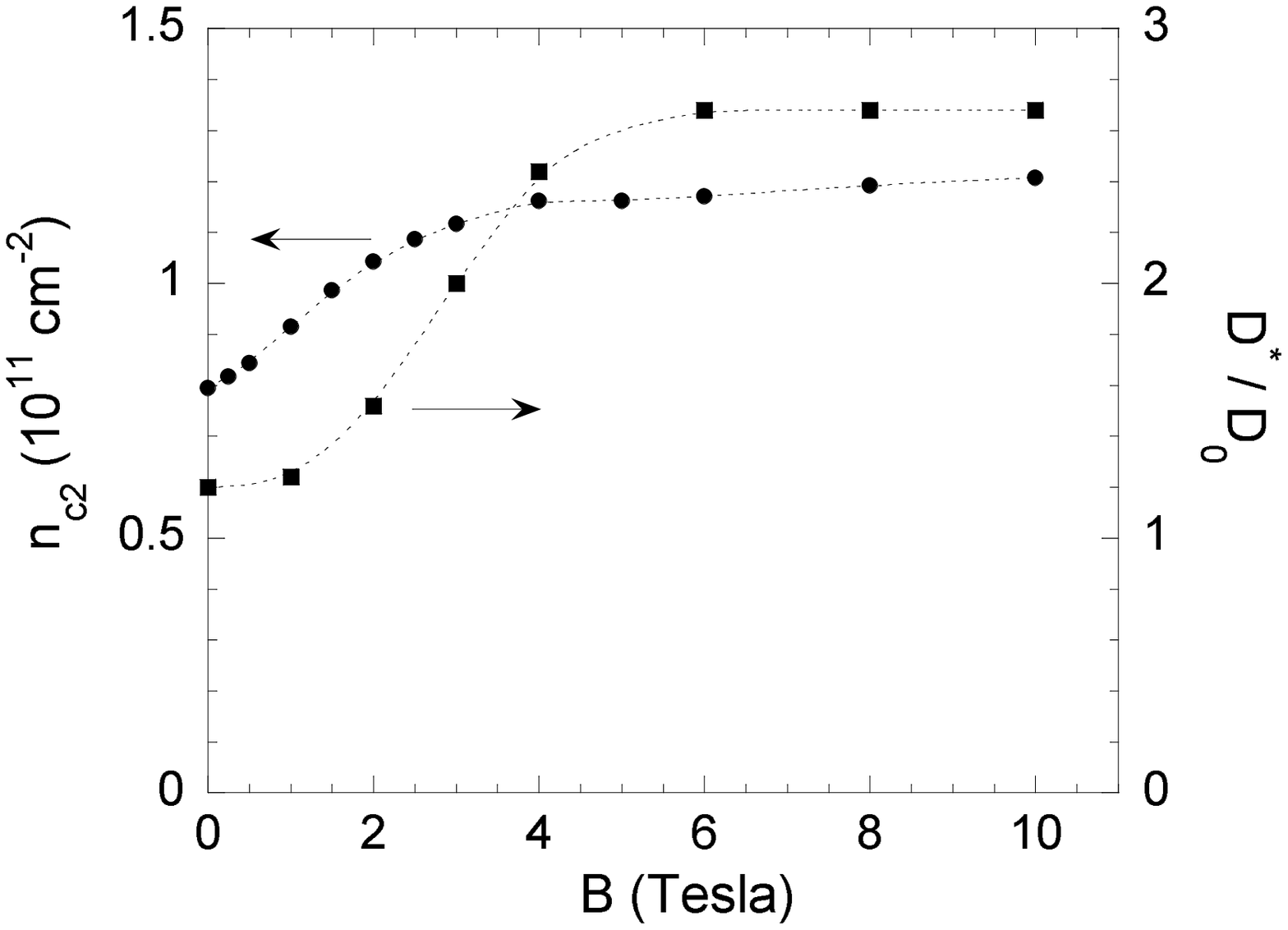,width=2.6in,bbllx=.5in,bblly=1.25in,bburx=7.25in,bbury=9.5in,angle=0}
}
\vspace{-.35in}
\hbox{
\hspace{-0.15in}
\refstepcounter{figure}
\parbox[b]{3.4in}{\baselineskip=12pt \egtrm FIG.~\thefigure.
Dependence of the critical electron density $n_{c2}$ (dots) and of
the thermodynamic density of states at the MIT (squares) on parallel
magnetic field. The dashed lines are guides to the
eye.\vspace{0.20in}
}
\label{4}
}
}
field, ``metallic'' and ``insulating'' $\rho(T)$ curves are
approximately symmetric on both sides of the transition (see above),
this symmetry completely disappears in a parallel magnetic field
({\it cf.} Figs.~\ref{1}(a) and \ref{1}(b,c)). An important
disagreement between theory \cite{GD} and experiment is the dramatic
weakening of the metallic temperature dependences in the magnetic
field, while theoretically, the derivative ${\rm d}\rho/{\rm d}T$ for
the spin-polarized state is expected to be twice as {\em high} for
the same electron density \cite{GD} (for more on this discrepancy,
see Ref.~\cite{mertes00}). Thus, the properties of the polarized
state cannot be deduced from those of the unpolarized state in a
straightforward way \cite{g}.

Another experimental fact that indicates the effect of polarization
is the change in the slope of the dependence of $E_a$ on $n_s$
(Fig.~\ref{3}) which we link, following Ref.~\cite{adkins76}, to the
thermodynamic density of states at the MIT. In zero magnetic field
the inverse slope, $D^*$, is close to the zero-field density of
states, $D_0$: $D^*\approx1.2\,D_0$ where $D_0=2m/\pi\hbar^2$,
$m=0.19\,m_e$, and $m_e$ is the free electron mass. In a parallel
magnetic field, $D^*$ increases and saturates at about $2.7\,D_0$
(see squares in Fig.~\ref{4}). This increase is quite surprising
since in the case of full spin polarization, one expects that the
density of states should {\it decrease} by a factor of two due to
lifting of the spin degeneracy \cite{rem1}. So, in many respects the
behavior of the spin-polarized electron system is peculiar.

In summary, we have compared two principal approaches for determining
the critical density for the metal-insulator transition in 2D. In
zero magnetic field, both definitions of the critical point are found
to be equivalent, strongly supporting the existence of a true $B=0$
MIT. With increasing parallel magnetic field, the $I-V$ criterion
gives the critical point which shifts to higher electron densities
and then saturates, which is likely to be a consequence of the spin
polarization of the 2D electrons. It is accompanied by the
disappearance/weakening of metallic temperature behavior of the
resistance so that the derivative criterion cannot be used. The fact
that the spin-polarized and spin-unpolarized dilute 2D electron
systems behave qualitatively differently poses important constraints
on the theory.

We gratefully acknowledge discussions with V.~T. Dolgopolov, D.
Heiman, M.~P. Sarachik, and S.~A. Vitkalov. This work was supported
by NSF grants DMR-9803440 and DMR-9988283, RFBR grant 01-02-16424,
and Sloan Foundation.





\end{multicols}

\begin{references}
\bibitem{away} Deeper in the insulating state, the density of states
decreases as compared to its value in the vicinity of the transition,
which results in an increase of the slope of the $E_a(n_s)$
dependence.
\bibitem{degeneracy} Obviously, account should be taken of the spin
degeneracy which is equal to $g_s=1$ ($g_s=2$) for the polarized
(unpolarized) state and involves parameters such as the Fermi energy
$E_F\propto 1/g_s$, the density of states $D\propto g_s$, etc.
\bibitem{rem1} We do not believe that the discrepancy is due to a
possible shift of the mobility edge when the electron density is
varied in the insulating phase. Firstly, for the small deviations of
$n_s$ from $n_{c2}$ in this experiment, the slope of $E_a$ {\it vs}
$n_s$ remains approximately constant, so that the screening ability
of the 2D system does not change with $n_s$, and therefore one
expects that the mobility edge does not shift. And secondly, no shift
of the mobility edge is seen in zero magnetic field because we and
others \cite{shashkin94,adkins76} have found $D^*\approx D_0$.
\end{references}
\end{document}